\documentclass[prl,twocolumn]{revtex4-2}

\usepackage{mathtools}
\usepackage{amsmath,bm}
\usepackage{amssymb}
\usepackage{subfigure}
\usepackage{setspace}
\usepackage{soul}
\usepackage{dcolumn}
\usepackage{color}

\usepackage[hidelinks]{hyperref}

\begin{document}

\title{Exploring Topology of 1D Quasiperiodic Metastructures through Modulated LEGO Resonators}

\author{Matheus I. N. Rosa, Yuning Guo, and Massimo Ruzzene}
\affiliation{P. Rady Department of Mechanical Engineering, University of Colorado Boulder, Boulder CO 80309}

\date{\today}

\begin{abstract}
We investigate the dynamics and topology of metastructures with quasiperiodically modulated local resonances. The concept is implemented on a LEGO beam featuring an array of tunable pillar-cone resonators. The versatility of the platform allows the experimental mapping of the beam's Hofstadter-like resonant spectrum, which is the first reported observation for an elastic medium. The non-trivial spectral gaps are classified by evaluating the integrated density of states of the bulk bands, which is experimentally verified through the observation of topological edge states localized at the boundaries. Results also show that the spatial location of the edge states can be varied through the selection of the phase of the resonator's modulation law. The presented results open new pathways for the design of metastructures with novel functionalities going beyond those encountered in periodic media by exploiting aperiodic patterning of local resonances and suggest a simple, viable platform for the observation of a variety of topological phenomena.
\end{abstract}

\maketitle

The discovery of topological insulators~\cite{hasan2010colloquium} has attracted significant interest from the metamaterials community due to promising prospects for robust wave localization and transport.~\cite{prodan2009topological,PhysRevLett.114.114301,wang2015topological,fleury2016floquet,lu2017observation,
mousavi2015topologically,susstrunk2015observation,PhysRevX.8.031074} Recent efforts focus on exploring higher dimensional topological effects in lower dimensional systems by exploiting virtual dimensions in parameter space.~\cite{qi2008topological,kraus2016quasiperiodicity,prodan2015virtual,
ozawa2016synthetic,lee2018electromagnetic} Indeed, edge states commonly attributed to the Quantum Hall Effect in 2D systems~\cite{wang2015topological} have been illustrated in 1D periodic~\cite{alvarez2019edge,rosa2019edge} and quasiperiodic~\cite{kraus2012topological,apigo2018topological,apigo2019observation,ni2019observation,Pal_2019,xia2020topological,gupta2020dynamics} systems, while 4D and 6D Quantum hall phases have been observed in 2D~\cite{zilberberg2018photonic,lohse2018exploring,rosa2020topological} and 3D~\cite{petrides2018six,lee2018electromagnetic} lattices. In addition to opening avenues for the exploration of novel topological wave physics phenomena, these investigations are also promising for technological applications and devices. For example, topological pumps as originally envisioned by Thouless~\cite{thouless1983quantization} were recently implemented,~\cite{kraus2012topological,rosa2019edge,riva2020edge,chen2020landau,nakajima2016topological,lohse2016thouless,grinberg2020robust,chen2019mechanical,brouzos2019non,longhi2019topological,PhysRevB.102.014305,xia2020experimental,cheng2020experimental} suggesting new mechanisms for robust energy transport in systems of a single spatial dimension. 

Among the many types of elastic metamaterials, locally resonant metastructures are particularly interesting due to the possibility of affecting dispersion at subwavelengths.~\cite{liu2000locally,yu2006flexural,sun2010theory,
oudich2011experimental,badreddine2012broadband,zhu2014chiral,
sugino2017general} Recent studies have explored the effects of aperiodicity and disorder~\cite{cardella2016manipulating,celli2019bandgap,beli2019wave,de2020experimental} for bandgap widening and producing rainbow effects. For example, elastic beams with arrays of identical resonators located according to quasi-periodic patterns investigated in~\cite{xia2020topological} were shown to feature additional spectral gaps hosting topological edge states. These were produced at no additional cost or increase in mass when compared to the nominal periodic configurations. Thus, quasi-periodic patterning of locally resonant metastructures may open new avenues for wave localization or attenuation in multiple bands, and for extending the behavior of periodic configurations.

\begin{figure}[t!]
\includegraphics[width=0.5\textwidth]{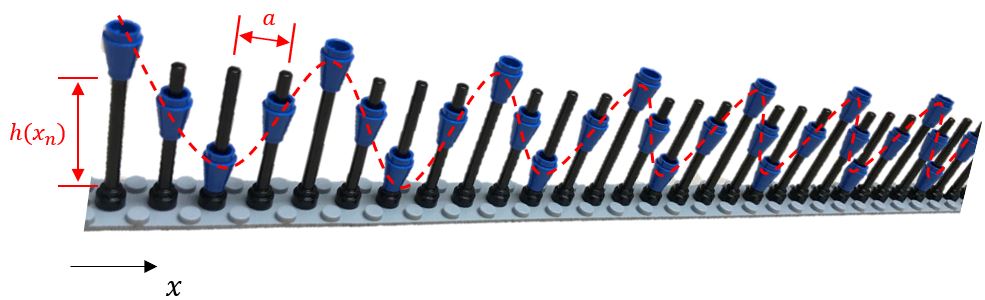}
\centering
\caption{Schematic of LEGO beam with pillar-cone resonators. A modulation of the cones' heights $h_n=h_0+\Delta h\sin(2\pi\theta n + \phi)$ is employed, represented by the dashed red line. The figure illustrates a periodic domain obtained with $\theta=1/4$, comprising 4 resonators per unit cell.}
	\label{Fig1}
\end{figure}

In this letter, we investigate locally resonant metastructures whose resonating attachments are tuned according to a quasi-periodic modulation law. We employ a LEGO elastic beam with pillar-cone resonators (Fig.~\ref{Fig1}), whose resonant frequencies are readily adjusted by sliding the cones along the pillars. LEGO bricks of this type were already employed in prior works to explore the effects of disorder in locally resonant metamaterials~\cite{celli2015manipulating,celli2019bandgap} through an experimental platform that is also suitable for investigations in the the context of quasi-periodic media. Indeed, the versatility of the platform enables the experimental mapping of the Hofstadter-like resonant spectrum of the beam, which to our knowledge is presented here for the first time for an elastic medium. Numerical simulations are conducted and allow for predictions of the spectral gaps, along with their topological classification based on the framework presented in~\cite{apigo2018topological}, while the experimental observation of an Hofstadter's spectrum in acoustic waveguides is reported in~\cite{ni2019observation}. Our experiments also illustrate the presence of topological edge states spanning the gaps, which are localized at one of the boundaries of the beam. In contrast to the investigations presented in~\cite{xia2020topological}, this work considers equally spaced resonators whose resonant frequencies are modulated instead of their spacing. This alternative approach may be advantageous especially for tunable devices, whereby the modulation of local properties such as the resonant frequency of piezoelectric shunt circuits~\cite{airoldi2011design,sugino2017investigation,marconi2020experimental} may be employed for versatile platforms without the need of physical reconfiguration. 

The considered elastic LEGO beam (gray solid in Fig.~\ref{Fig1}) is equipped with an array of resonators of equal spacing $a$, whose resonance frequencies are modified by sliding the cones (blue) along the pillars (black). The height $h_n$ of the cone in the $n$-th resonator is assigned according to the law
\begin{equation}\label{modEQ}
h_n=h_0 + \Delta h\sin(2\pi\theta n + \phi),
\end{equation}
where $h_0$ and $\Delta h$ denote the offset and amplitude of the modulation. Such modulation can be visualized as the sampling of a sinusoidal waveform $h(x)=h_0+\Delta h\sin(2\pi \theta x + \phi)$ (dashed red line in Fig.~\ref{Fig1}) at locations $x_n=n$~\cite{rosa2019edge}. Alternatively, the law can also be visualized as the projection from an array of circles.~\cite{apigo2019observation,Pal_2019,xia2020topological} The parameter $\theta$ controls the periodicity of the modulation: rational $\theta$ values of the form $p/q$ with co-prime $p,q$ identify periodic structures with $q$ resonators per unit cell, while irrational $\theta$ values are associated with quasiperiodic domains. The illustration in Fig.~\ref{Fig1} exemplifies a periodic domain with $\theta=1/4$, comprising 4 resonators per unit cell. The phase (or phason) $\phi$ does not affect the periodicity of the domain, but is a parameter which reveals the existence of edge states and defines their localization at one of the two boundaries.~\cite{rosa2019edge,xia2020experimental,cheng2020experimental}

\begin{figure*}[t!]
\centering
\subfigure[]{\includegraphics[height=0.35\textwidth]{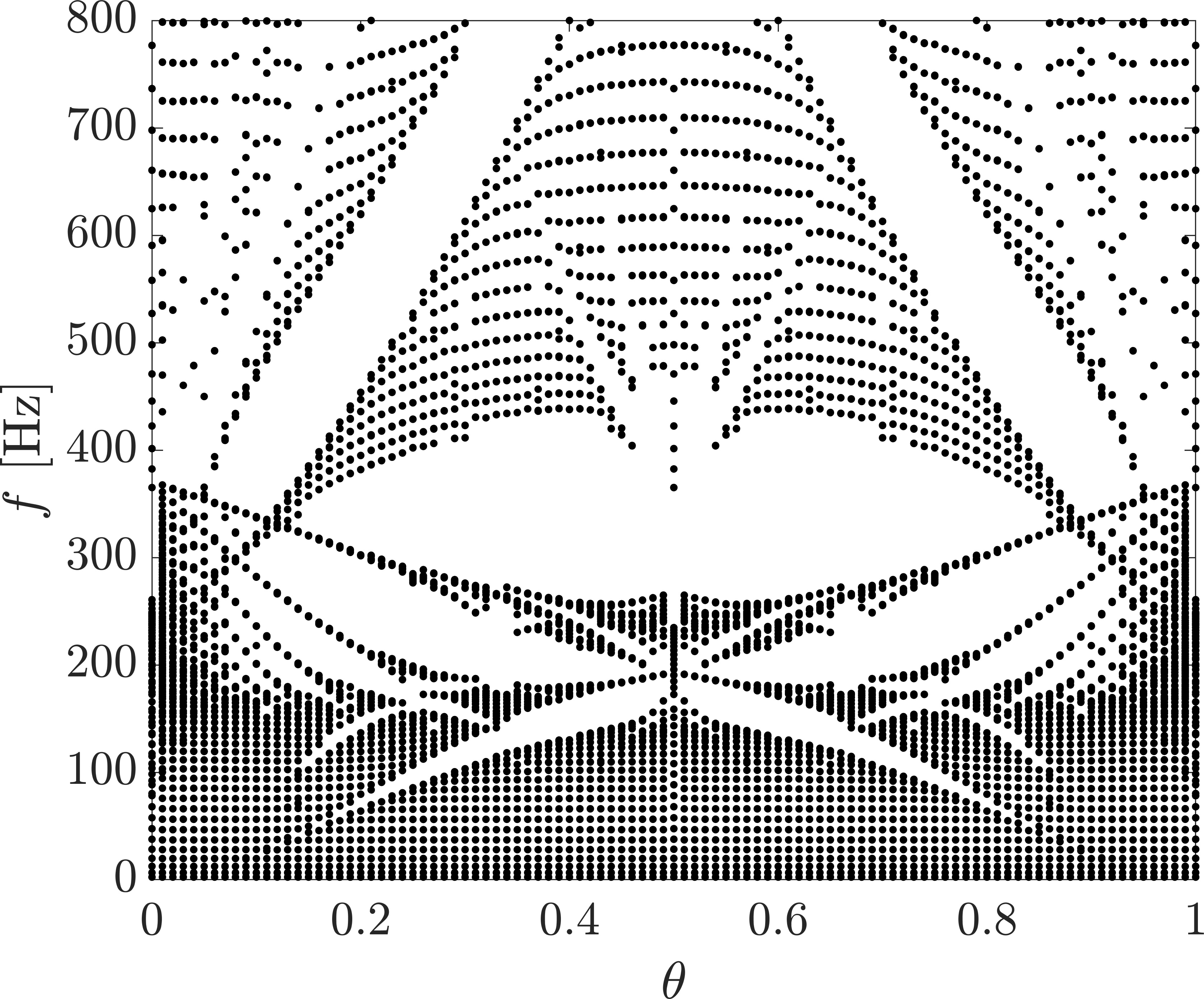}\label{Fig2a}}\hspace{5mm}
\subfigure[]{\includegraphics[height=0.35\textwidth]{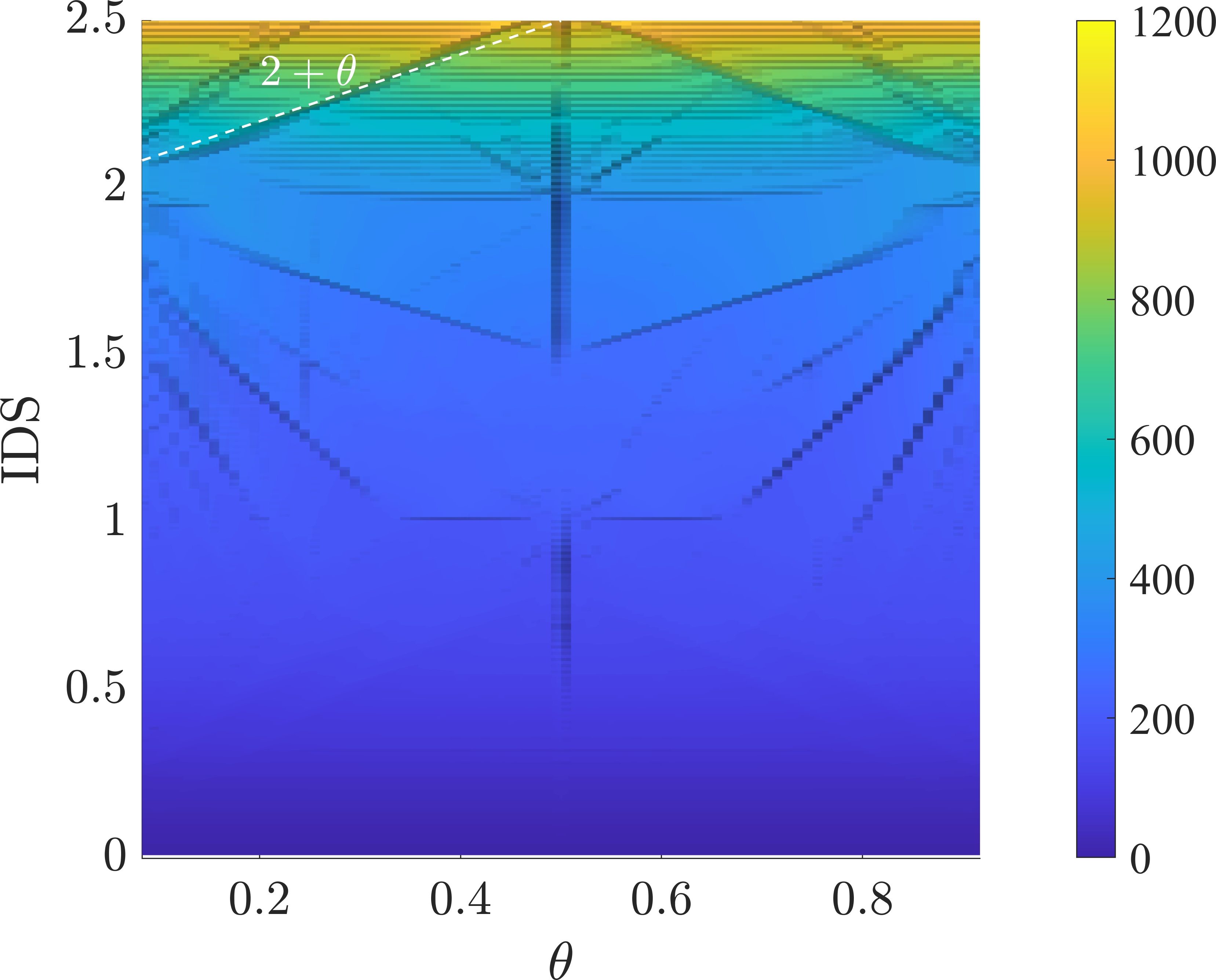}\label{Fig2b}}
\caption{Spectrum (a) and IDS (b) of quasiperiodic LEGO beam as a function of $\theta$. The non-trivial spectral gaps in (a) are associated with non-horizontal straight lines in (b). }
	\label{Fig2}
\end{figure*}

The spectral properties of the beam are first characterized for the case of uniform distribution of resonators ($\theta=0$) by conducting 3D finite element (FE) simulations and subequent experimental verification. Following the process detailed in the Supplemental Materials (SM)~\cite{SM}, the local resonant gap produced by the pillar-cone resonator is mapped as a function of the height of the cone $h_0$. This shows that the center frequency of the gap varies approximately from $200$Hz to $350$Hz as $h_0$ is varied from $0$ to $30$mm (other relevant physical parameters are provided in \cite{SM}). The spectrum for the quasi-periodic modulation (Eqn.~\ref{modEQ}) is then conveniently mapped by using periodic approximants~\cite{apigo2018topological,Pal_2019}. To this end, the eigenfrequencies of a finite beam comprising $N=100$ resonators are computed by applying periodic boundary conditions for $\theta$ varying in steps of $1/100$, corresponding to the subset of periodic values for the chosen structure size. The modulation parameters $h_0,\Delta h=15$mm are considered to explore the entire range of height variation (from $0$ to $30$mm). Results reported in Fig.~\ref{Fig2a} show the Hofstadter-like spectrum of the beam as a function of $\theta$. For $\theta=0$, a single local resonant gap exists in the $f=\{280,360\}$Hz range, which corresponds to the gap predicted by the analysis of uniform resonators with $h_0=15$mm~\cite{SM}. Interestingly, such local resonant gap is quickly transformed into a series of additional gaps at lower and higher frequencies as $\theta$ is varied through the reconfiguration of the cones' heights defined by the modulation law of Eqn.~\ref{modEQ}. We note that in the analysis we only include modes with a significant component of motion that is perpendicular to the beam axis (bending modes). These are separated from the other polarizations by considering a polarization factor that filters out predominantly longitudinal or torsional modes. Definition of the polarization factor, and of the modal filtering process are found in~\cite{SM}

The topological properties of the spectrum are revealed by computing the integrated density of states (IDS)~\cite{apigo2018topological,Pal_2019}, which is displayed in Fig.~\ref{Fig2b}. The rendering of the IDS highlights straight lines which are associated with the spectral gaps. Non-horizontal lines indicate non-trivial gaps, identified by a nonzero Chern numbers that are evaluated from the slope of the corresponding gap line.~\cite{apigo2018topological,ni2019observation} The most prominant gap in Fig.~\ref{Fig2a} (rising up from 400-800 Hz approximately) is labeled by the fitting highlighted in Fig.~\ref{Fig2b} (white dashed line), illustrating that $IDS=2+\theta$ for that gap, and thus its Chern number is $C=1$.

\begin{figure*}
\centering
\subfigure[]{\includegraphics[height=0.35\textwidth]{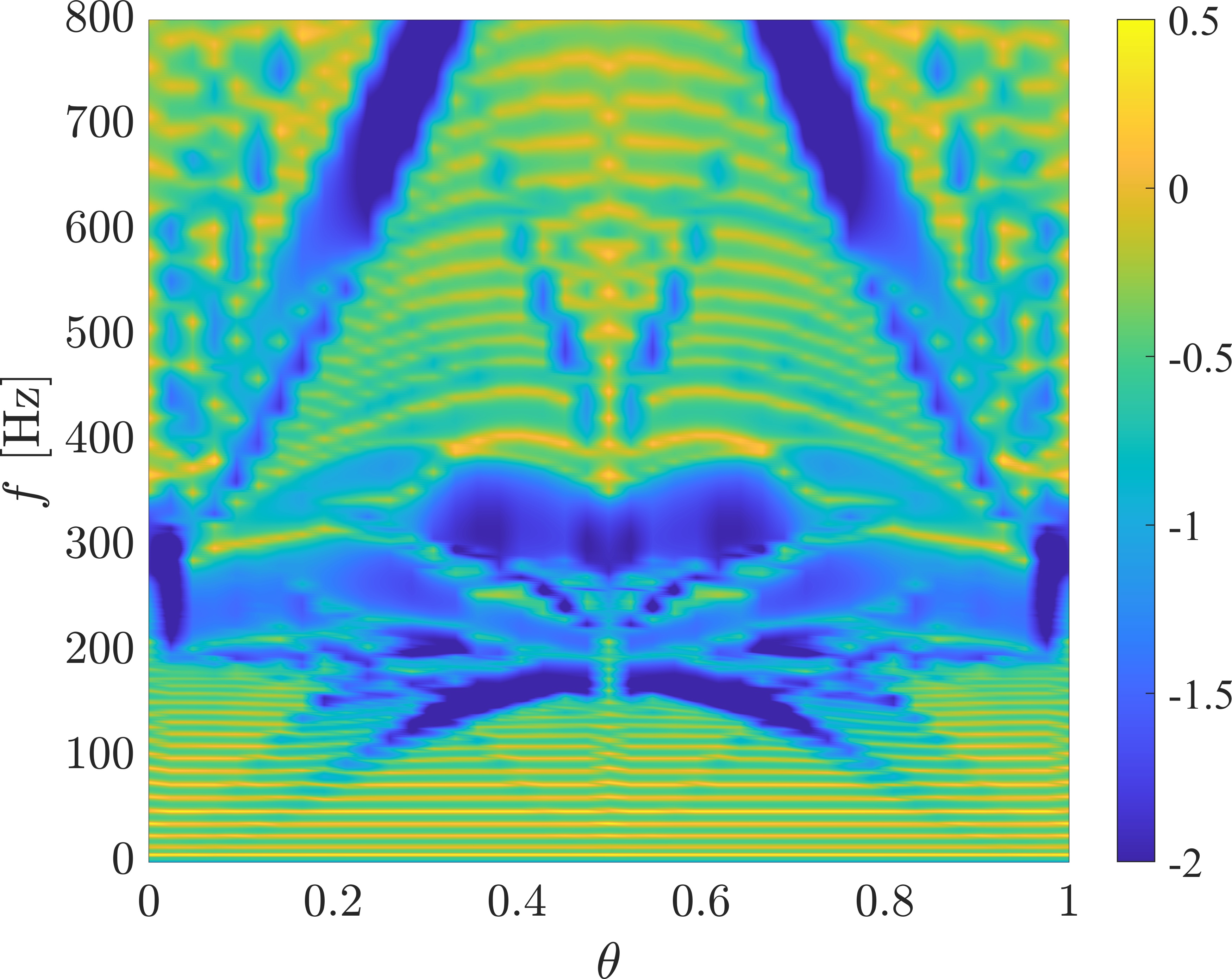}\label{Fig3a}}\hspace{5mm}
\subfigure[]{\includegraphics[height=0.35\textwidth]{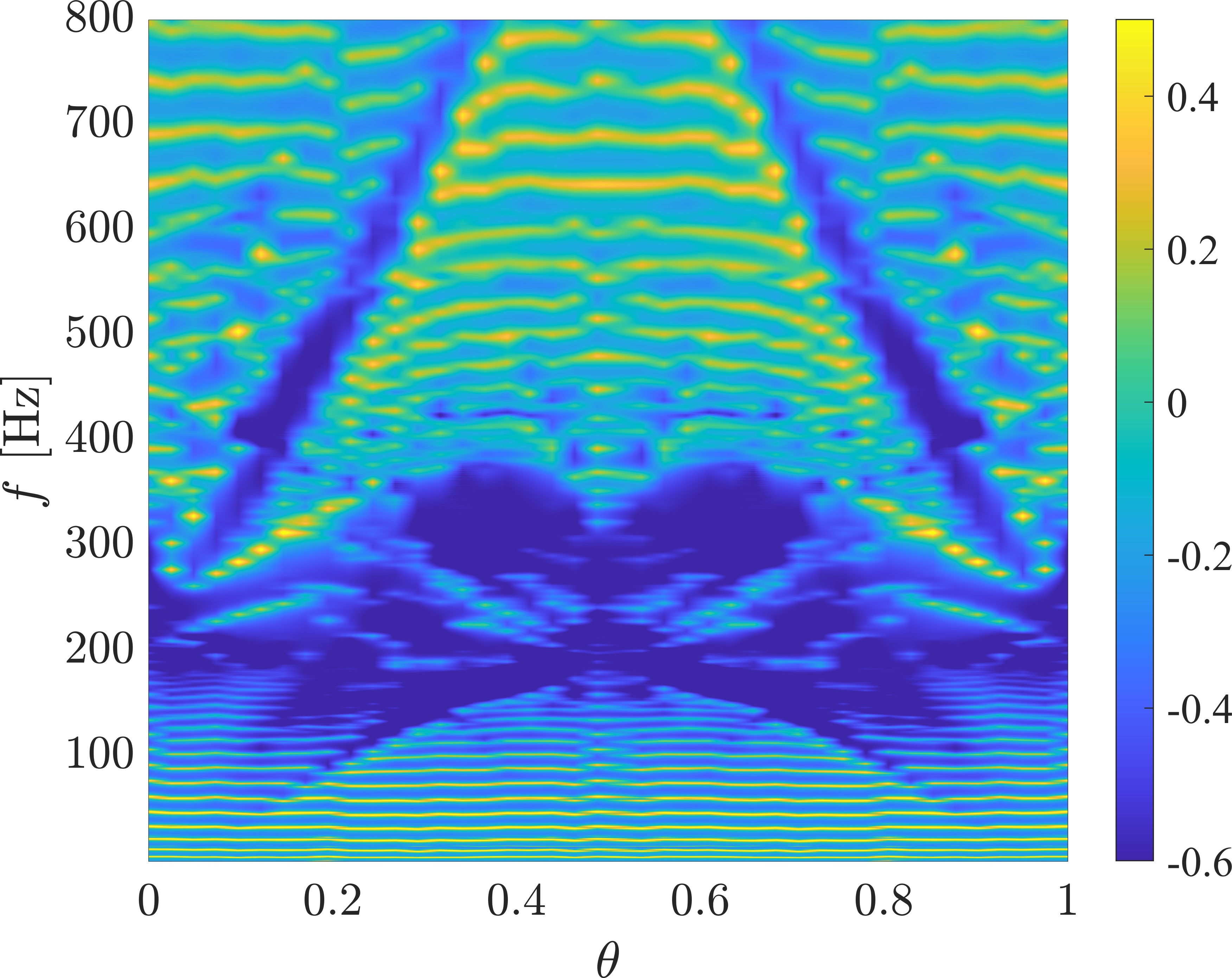}\label{Fig3b}}
\caption{Numerical (a) and experimental (b) average transmission of quasiperiodic beam as a function of $\theta$, capturing the features of the Hofstadter-like spectrum.}
	\label{Fig3}
\end{figure*}

The experimental investigations have as a first goal the mapping of the Hofstadter-like spectrum of Fig.~\ref{Fig2a}. A beam comprising $N=42$ resonators is clamped at the right end and excited at the left end by an electrodynamic shaker~\cite{SM}. A broadband pseudo-random signal in the range $f=\{0, 800\}$Hz is applied to excite the bending motion of the beam. The motion is recorded by a scanning laser doppler vibrometer (SLDV) at a total of 80 points aligned along the span of the beam. Finally, the transmission is calculated by computing the ratio between measured velocity at the measurement points to the beam velocity at the location of shaker, i.e. the input point. Results presented in Fig.~\ref{Fig3} compare the simulation results computed via 3D FE (a) with experimental measurements (b). A total of $20$ experiments are conducted for $\theta$ varying from $0$ to $0.5$, and results are presented in the range $\theta \in [0,1]$ for better visualization, which utilizes symmetry in the quasi-periodic pattern allowing mirroring the results obtained for $\theta \in [0,0.5]$. In the figure, the color denotes the magnitude of the log scale of the transmission amplitude averaged across all measurement points. The experimental results in (b) are in good agreement with the numerical results in (a), and overall display most of the features predicted by the spectrum of Fig.~\ref{Fig2a}, confirming in particular the presence of the largest gap labeled with $C=1$. 

\begin{figure*}
\centering
\subfigure[]{\includegraphics[height=0.27\textwidth]{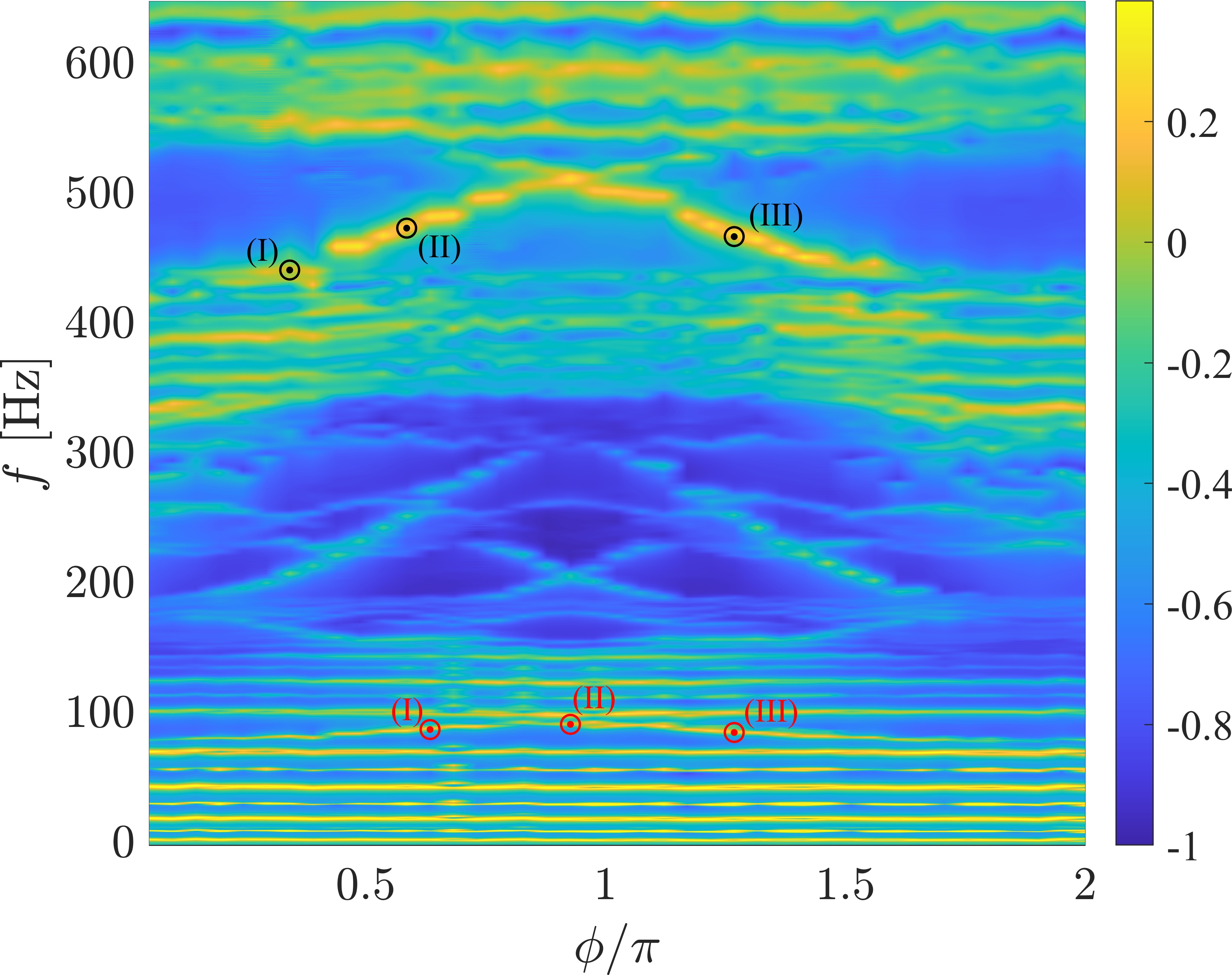}\label{Fig4a}}
\subfigure[]{\includegraphics[height=0.26\textwidth]{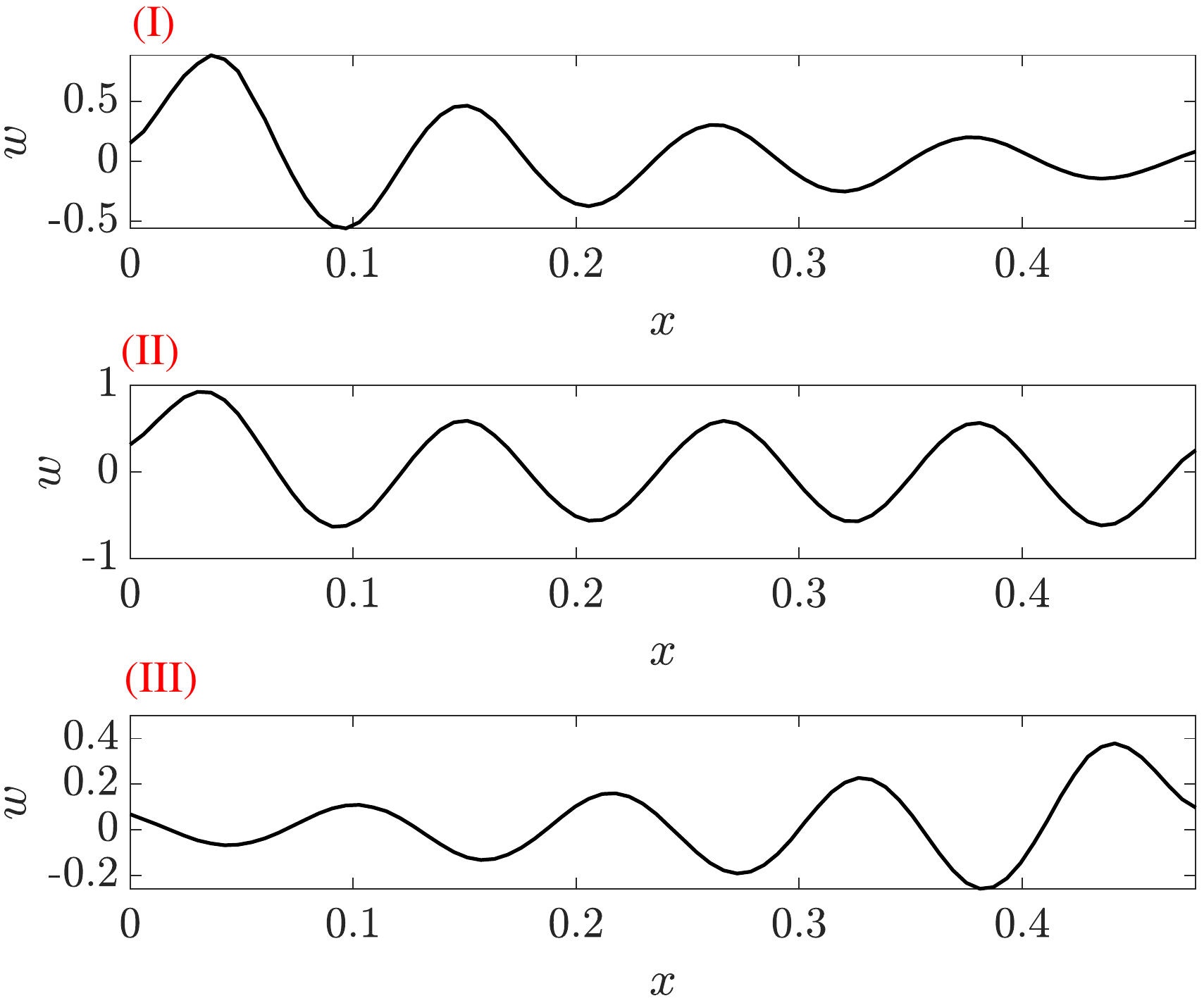}\label{Fig4b}}
\subfigure[]{\includegraphics[height=0.26\textwidth]{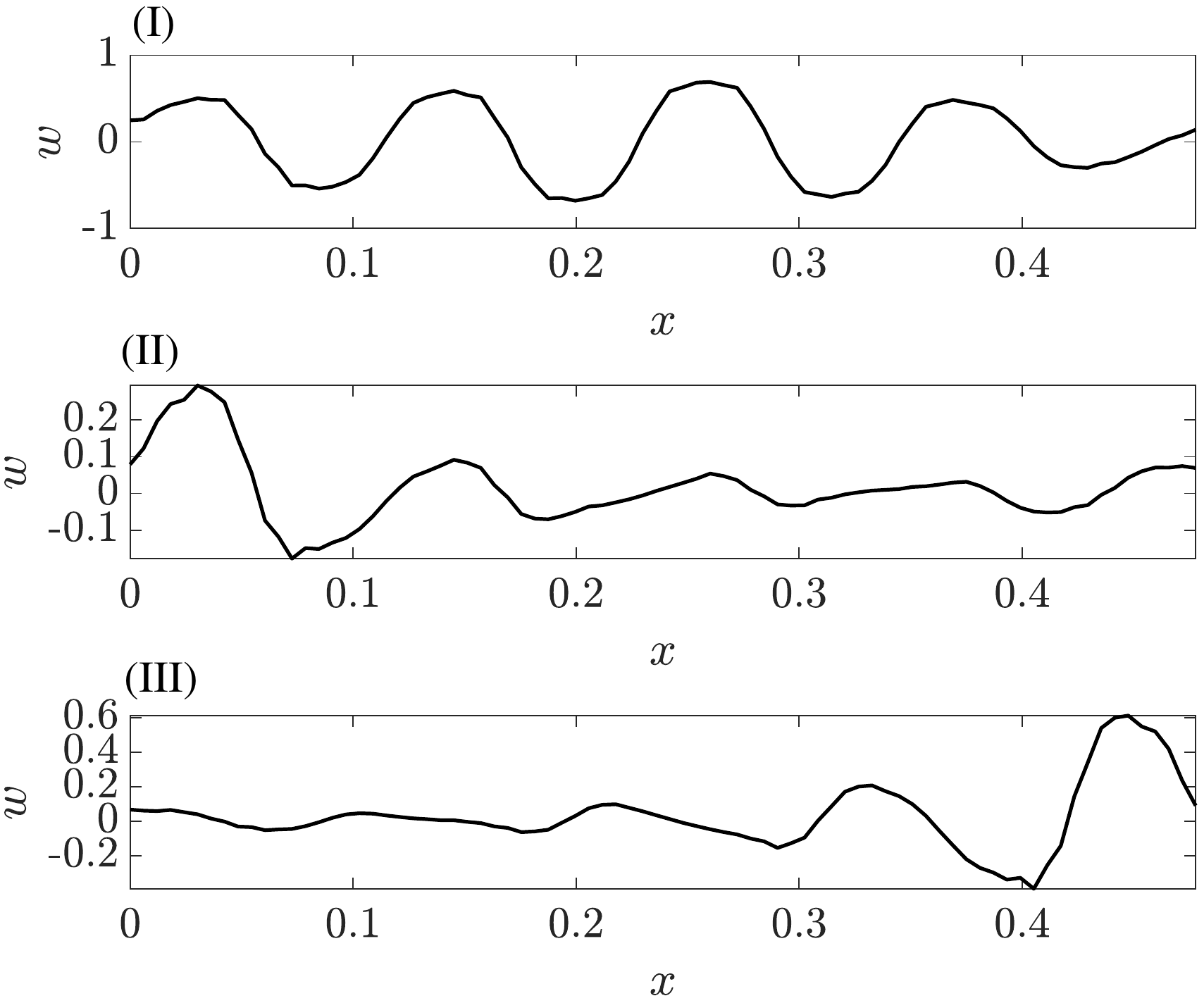}\label{Fig4c}}
\caption{Experimentally measured average transmission of beam with $\theta=0.2$ as a function of $\phi$. Edge states are observed to span the non-trivial gaps. Representative modes marked in (a) are displayed in (b), illustrating the edge states and their transitions.}
\label{Fig4}
\end{figure*}

Next, we experimentally demonstrate the existence of topological edge states spanning the non-trivial gaps. We consider a representative case of $\theta=0.2$, corresponding to a periodic beam with $5$ resonators per unit cell, and explore the spectral variation with the phason parameter $\phi$ in Eqn.~\eqref{modEQ}. Two sets of experiments are conducted on the beam with $N=42$ resonators in the same conditions as previously described, but with excitation at the left boundary in one case, and at the right boundary in the other. For each case, 41 experiments are conducted for $\phi$ varying within the $\{0, 2\pi \}$ range. As illustrated in the results detailed in the SM~\cite{SM}, the spectrum averaged across all points along the beam captures one group of branches corresponding to the edge states. The left-localized branches are captured by the left-excitation experiment, while the right-localized branches are captured by the right-excitation experiment. We here provide a single spectral characterization of the beam in Fig.~\ref{Fig4a} obtained by combining the results from both experiments, which allows the observation of both left- and right-localized branches of the edge states. Two sets of experimental modes illustrating a transitions of the edge states are marked in Fig.~\ref{Fig4a} and displayed in Figs.~\ref{Fig4}(b,c). In the first case (Fig.~\ref{Fig4b}), a smooth transition from left-localized (I), to bulk mode (II), and finally to right-localized (III) is observed. In the second case (Fig.~\ref{Fig4c}), the branches of the edge state cross, and a transition from bulk (I) to left-localized (II) is first illustrated, while a representative right-localized response (III) from the right-localized branch is also displayed. These transitions are commonly explored for topological pumping in passive systems exploiting an extra spatial dimension~\cite{kraus2012topological,rosa2019edge,riva2020edge,chen2020landau}, or in active systems by modulating the phason in time $\phi$~\cite{xia2020experimental,cheng2020experimental}. Video animations of these experimental vibration modes captured by the SLDV are provided in the SM.~\cite{SM}

The results presented in this letter illustrate experimentally how quasi-periodic patterning of the resonant attachments on 1D metastructures can be used to open additional non-trivial gaps hosting topological edge states. Such analysis expands the results reported in~\cite{xia2020topological} by considering modulations of the resonant frequencies instead of operating on the locations of inclusions. This provides potential for implementation of these patterns using tunable devices, such as electromechanical waveguides~\cite{xia2020experimental}. In addition, the LEGO platform is shown to be suitable for the exploration of elastic wave phenomena, and may applicable for other related explorations. Future work may focus on extending the presented analysis to two-dimensional (2D) quasiperiodic media, where higher dimensional topologies such as the 4D Quantum Hall Effect may be explored~\cite{lohse2018exploring,rosa2020topological,cheng2020mapping}.

\begin{acknowledgments}
The authors gratefully acknowledge the support from the National Science Foundation (NSF) through the EFRI 1741685 grant and from the Army Research office through grant W911NF-18-1-0036. The data that support the findings of this study are available from the corresponding author upon reasonable request.
\end{acknowledgments}

\bibliography{References}
\end{document}